\documentclass[12pt]{article}
\makeatletter

\@addtoreset{equation}{section}
\makeatother

\def\N{{\cal N}}

\def\O{{\cal O}}

\def\cob{\delta}

\def\Tr{{\rm Tr}}

\def\hf{{1\over 2}}

\def\R{{\bf R}}
\def\o{\over}

\def\del{\partial}

\def\bra{\langle}
\def\ket{\rangle}
\def\lf{\left}
\def\ri{\right}
\def\riya{\rightarrow}

\def\h#1{\widehat{#1}}

\def\bt{\beta}

\def\al{\alpha}
\def\om{\omega}

\def\tens{\otimes}

\def\dag{\dagger}
\def\rt#1{\sqrt{#1}}

\def\st{\star}

\def\sitarel#1#2{\mathrel{\mathop{\kern0pt #1}\limits_{#2}}}

\def\a{a^{\dag}}

\def\sd{s^{\dagger}}
\newcommand{\nn}{\nonumber \\}
\newcommand{\kket}{\rangle\rangle}
\newcommand{\bbra}{\langle\langle}

\begin{document}
                        
\rightline{\vbox{\hbox{KEK-TH-777}\hbox{hep-th/0107101}}}
\baselineskip 0.8cm
\vspace*{2.5cm}
\begin{center}
{\LARGE\bf Comma Vertex and String Field Algebra}
\end{center}
\vskip 8ex
\begin{center}
  Kazuyuki Furuuchi and Kazumi Okuyama\\
\vskip 2mm
  {\sl Theory Group, KEK, Tsukuba, Ibaraki 305-0801, Japan} 
\vskip 2mm
  {\tt furuuchi@post.kek.jp, kazumi@post.kek.jp}
\end{center}
\vskip 25mm
%
\baselineskip=6mm
\centerline{{\bf Abstract}}
\vskip 5mm
We study the matter part of the algebra of open string fields
using the 3-string vertex over the sliver state, which we call
``comma vertex''. By generalizing this comma vertex
to the $N$-string overlap, we obtain a closed form of
the Neumann coefficients in the $N$-string vertex and
discuss its relation to the oscillator representation of
wedge states.

\vskip 50mm
\noindent
July 2001
\newpage

\section{Introduction}
Recently, the cubic open string field theory \cite{Witten}
around the tachyon vacuum,
or the ``vacuum string field theory'' \cite{VSFT,VSFTreview} was proposed.
This theory contains no physical open string excitation
and is conjectured to describe the appearance of closed strings after 
the open string tachyon condensation.  
The matter part of the 
solution of the classical equation of motion around this vacuum is
given by projection elements in the star algebra of 
open string fields \cite{RSZ1}.

As an example of projection, 
there is a state called sliver \cite{RZ},
which is special in a sense
that this state can be defined in an arbitrary boundary CFT.
In the case of flat background
a projection was also constructed 
in the oscillator formalism \cite{KosteleckyPotting} 
and was identified with the sliver with numerical evidence \cite{RSZ1}.
To construct projections more systematically, 
it is important to find a matrix 
representation of open string field \cite{RSZ2,GrossTaylor,KawanoOkuyama}
(see \cite{RSZ3,Matsuo,David,Bars} for related issues).
The essential ingredient behind this matrix representation
is the so-called ``comma'' form of the overlap vertex 
\cite{Bordes,Abdurrahman,ChanTsou}. Interestingly, if we perform a Bogoliubov 
transformation to make the sliver state 
to be a new Fock vacuum, the 3-string vertex 
takes a comma form \cite{KosteleckyPotting}.
In this respect the sliver is a
key to the structure of the string field algebra.
In this paper, we study the properties of the overlap vertices constructed 
over the sliver state, and discuss some applications.  

This paper is organized as follows: In section 2, we consider the 3-string
vertex over the sliver state. In section 3, we analyze the structure 
of string field algebra using this vertex. In section 4, we construct the
oscillator representation of wedge states and find a closed form of 
the Neumann coefficient of $N$-string vertex. Section 5 is devoted to 
the discussions.

\section{Comma Vertex}
In this section, we review the 3-string vertex constructed over the
sliver state \cite{KosteleckyPotting}. In the following,
we will call this vertex the ``comma vertex'' 
\cite{Bordes,Abdurrahman,ChanTsou}.  

\subsection{3-string Neumann coefficient}

The matter part of the 
3-string vertex in zero-momentum sector is given by
\cite{GrossJevicki,Samuel,Cremmer}
\begin{equation}
 |V_3\ket={\cal N}_{V_3}^{26}
\exp\lf(-\hf\sum_{\stackrel{r,s=1,2,3}{m,n\geq 1}} 
a^{r\dagger}_m V_{3\,mn}^{rs} a_n^{s\dagger} \ri)
|0\ket_{123}.
\end{equation}
${\cal N}_{V_3}$ is a normalization constant.
In this paper we concentrate
on zero-momentum sector 
but algebraic relations we will obtain are the same
if we consider whole sector
\cite{KosteleckyPotting,RSZ1,GrossTaylor}.
The identity string field $|I\ket $ and the reflector $\bra R(1,2) |$
is given by (up to normalization)
\begin{eqnarray}
 \label{IRa}
|I\ket = e^{-\frac{1}{2} a^{\dagger}Ca^{\dagger}} |0\ket, \quad
\bra R(1,2)| =
{}_{12}\bra 0 | e^{-a_1Ca_2},
\end{eqnarray}
where $C_{nm}=(-1)^n\cob_{nm}$.
We define star product of string fields as 
\begin{equation}
 |A\st B\ket_1={}_3\bra r(A)|{}_2\bra r(B)||V_3\ket_{123},
\end{equation}
where
\begin{equation}
 {}_1\bra r(A)|=\bra R(1,2)|A\ket_2.
\end{equation}

As preliminaries, we list some useful formulas satisfied by
the Neumann coefficients $V^{rs}_3$ in the 3-string vertex.
We introduce the matrices $X,Y,Z$ by 
\begin{equation}
X=CV^{11}_3,\quad Y=CV^{12}_3,\quad Z=CV^{21}_3.
\end{equation}
It can be shown that these matrices commute with each other
and satisfy the following relations:
\begin{eqnarray}
 &&X+Y+Z=1, \\
&&X^2+Y^2+Z^2=1,\quad YZ=X^2-X,\\
&&Y^3+Z^3=2X^3-3X^2+1, \\
&&CXC=X,\quad CYC =Z.
\end{eqnarray}
Using these matrices,
one can construct the orthogonal projections $L$ and $R$ 
\footnote{In \cite{RSZ2}, $L$ and $R$ were denoted 
$\rho_1$ and $\rho_2$, respectively.}
\begin{equation}
 \label{LR}
L={Y(1-TX)+TZ^2\o (1+T)(1-X)},\quad 
R={Z(1-TX)+TY^2\o (1+T)(1-X)},  
\end{equation}
which satisfy
\begin{equation}
 \label{eqLR}
L+R=1,\quad L^2=L,\quad R^2=R,\quad LR=0,\quad L=CRC.
\end{equation}
Here $T$ is defined by 
\begin{eqnarray}
T=\frac{1}{2X} \lf(1+X - \sqrt{(1-X)(1+3X)}\ri),
\end{eqnarray}
and satisfies
\begin{equation}
 \label{eqT}
XT^2-(1+X)T+X=0.
\end{equation}
This matrix $T$ appears in the sliver state 
\cite{KosteleckyPotting,RSZ1}
\begin{equation}
 |\Xi\ket={\cal N}_{\infty}^{26}\exp\lf(-\hf\a CT\a\ri)|0\ket ,
\end{equation}
which satisfies $|\Xi\ket \star |\Xi\ket = |\Xi \ket$.
The matrices $Y$ and $Z$  are decomposed into $L$ and $R$ as
\begin{eqnarray}
Y&=&L(1-TX)+RX(T-1)=(1-TX)(L-TR), \\
Z&=&LX(T-1)+R(1-TX)=(1-TX)(R-TL).
\end{eqnarray}

\subsection{Vertex over the Sliver}
We introduce new oscillators $s$ which annihilate the sliver
by performing a Bogoliubov transformation
from the original oscillators $a$
\begin{equation}
 \label{Bogo}
s={1\o\rt{1-T^2}}(a+CT\a),\quad \sd={1\o\rt{1-T^2}}(\a+CTa),
\end{equation}
\begin{equation}
s|\Xi\ket=0.
\end{equation}
The 3-string vertex in this new basis (up to normalization) turns out to be 
\cite{KosteleckyPotting} 
\begin{equation}
|V_3\ket=\exp\lf(-\sum_{r=1}^3\sd_rCL\sd_{r+1}\ri)|\Xi\ket_{123}=
\exp\lf(-\hf \sum_{r,s=1}^3\sd_r\h{V}^{rs}_3\sd_s\ri)|\Xi\ket_{123},
\label{comma3}
\end{equation}
where $\h{V}^{rs}_3$ is given by 
\begin{equation}
 \label{hV}
C\h{V}_3=(1-CV_3T)^{-1}(CV_3-T)=\lf(\matrix{0&L&R\cr R&0&L\cr L&R&0}\ri).
\end{equation}
The identity string field and the reflector in this new basis
are found to be
\begin{equation}
 \label{IRs}
|I\ket=e^{-\frac{1}{2} s^\dagger C s^\dagger}|\Xi\ket,\quad 
\bra R(1,2)|={}_{12}\bra\Xi|e^{-s_1 C s_2 }. 
\end{equation}
Using the projections $L$ and $R$, we can define the splitting of $s$
into two parts
\begin{equation}
 s=\sum_{\al}s_{L\al}e_{\al}+s_{R\al}f_{\al},
\end{equation}
where $e_{\al}$ and $f_{\al}$ are the orthonormal basis of the eigenspace
of $L$ and $R$:
\begin{equation}
 e_{\al}\cdot e_{\bt}=f_{\al}\cdot f_{\bt}=\cob_{\al\bt},\quad
e_{\al}\cdot f_{\bt}=0,\quad f_{\al}=-Ce_{\al},
\end{equation}
\begin{equation}
 Le_{\al}=e_{\al},\quad Re_{\al}=0,\quad Lf_{\al}=0,\quad Rf_{\al}=f_{\al},
\end{equation}
\begin{equation}
 L=\sum_{\al}e_{\al}e_{\al}^T,\quad R=\sum_{\al}f_{\al}f_{\al}^T.
\end{equation}
Since $s_L$ and $s_R$ commute,
\begin{equation}
 [s_{L\al},\sd_{L\bt}]=[s_{R\al},\sd_{R\bt}]=\cob_{\al\bt},\quad
[s_{L\al},\sd_{R\bt}]=0,
\end{equation}
the Hilbert space factorizes into the Fock spaces of $s_L$ and $s_R$
\begin{equation}
{\cal A}_{str}={\cal H}_L\tens {\cal H}_R.
\end{equation}
In terms of $s_{L,R}$, the 3-string vertex takes a simple
form (``comma'' form)
\begin{equation}
|V_3\ket=\exp\Big(\sd_{1R}\sd_{2L}+\sd_{2R}\sd_{3L}+\sd_{3R}\sd_{1L}
\Big)|\Xi\ket_{123} ,
\end{equation}
where
\begin{equation}
 \sd_L\sd_R=\sum_{\al}\sd_{L\al}\sd_{R\al}.
\end{equation}
The identity string field and the reflector become
\begin{equation}
|I\ket=e^{\sd_R\sd_L}|\Xi\ket,\quad 
\bra R(1,2)|={}_{12}\bra\Xi|e^{s_{1R}s_{2L}+s_{2R}s_{1L}}. 
\end{equation}

\subsection{On the Structure of the Vertex over the Sliver}

The vanishing of the diagonal components of $\h{V}_3$
in (\ref{hV}) is simply explained from the
identity
$|\Xi\ket \star |\Xi\ket = |\Xi\ket$ 
written in terms of the
$s$-oscillators
\cite{KosteleckyPotting}.
The identities 
$LR = 0$ and $RL = 0$,
which seem somewhat accidental from the complicated
expression (\ref{LR}),
also simply follow from
$|\Xi\ket \star | I \ket = |\Xi\ket$ and
$| I \ket \star | \Xi \ket = |\Xi\ket$ 
in the $s$-oscillator basis:
\begin{eqnarray}
 \label{XiI}
\lf(|\Xi\ket \star |I\ket \ri)_1
&=&
{}_3\bra \Xi |{}_2\bra I | | V_3 \ket_{123} \nn
&=&
{}_3\bra \Xi |{}_2\bra \Xi|\exp\lf(-\frac{1}{2}s_2 C s_2 \ri)
\exp\lf(-\hf \sum_{r,s=1}^3\sd_r\h{V}^{rs}_3\sd_s\ri) 
|\Xi\ket_{123} \nn
&=&
{}_2\bra \Xi|
\exp\lf(-\frac{1}{2}s_2 C s_2 \ri)
\exp\lf(-s_1^{\dagger}\h{V}^{12}s_2^\dagger\ri) |\Xi\ket_{12} \nn
&=&\exp\lf(-\hf s_1^\dagger \h{V}^{12} C \h{V}^{21} s_1^{\dagger} \ri)
|\Xi\ket_1 ,
\end{eqnarray}
where use has been made for the formula
\begin{eqnarray}
&&\bra 0| 
\exp\lf(\lambda_i a_i -\frac{1}{2} P_{ij} a_i a_j \ri)
\exp\lf(\mu_i a_i^{\dagger} -\frac{1}{2} Q_{ij} a_i^{\dagger} a_j^{\dagger}\ri)
| 0 \ket \nn
&&\quad=
\det K^{-1/2}\,
\exp\lf(\mu K^{-1}\lambda -\frac{1}{2} \lambda Q K^{-1} \lambda
-\frac{1}{2} \mu  K^{-1} P \mu \ri),\quad
K \equiv 1 - PQ.  
\end{eqnarray}
From eq.(\ref{XiI}) we observe
$|\Xi\ket \star |I\ket = |\Xi\ket $ implies
$\h{V}^{12} C \h{V}^{21} = 0$, which is equivalent to 
$LR = 0$.
Similarly, $RL=0$ can be understood from $|I\ket \star |\Xi\ket = |\Xi\ket $.

We briefly comment on other identities in (\ref{eqLR}).
$\h{V}_3$ can be diagonalized in the $r,s$ indices
by a matrix $\O$ satisfying $\O^{-1} = \O^{\dagger}$:
\begin{eqnarray}
 \label{Odiag}
\h{V}_3 = \O^{-1} \h{V}_D \O,
\end{eqnarray}
with
\begin{eqnarray}
 \label{Vdiag}
\h{V}_D \equiv
\lf(\matrix{C&0&0\cr 0&\Lambda&0\cr 0&0&\bar{\Lambda}}\ri),
\quad
\O = 
\frac{1}{\sqrt{3}}
\lf(\matrix{1&1&1\cr \omega&\bar{\omega}&1\cr \bar{\omega}&\omega&1}\ri).
\end{eqnarray}
Here $\omega \equiv \exp\lf(2\pi i/3 \ri)$.
$\h{V}_D^{11} = C$ follows because
this component represents identity string field type 
overlap condition (see (\ref{IRa}) and (\ref{IRs})).
From (\ref{Odiag}) and (\ref{Vdiag}), we obtain
\begin{eqnarray}
C &=& \h{V}_3^{12} + \h{V}_3^{21}, \nn
\Lambda &=& \bar{\omega} \h{V}_3^{12} + \omega \h{V}_3^{21}, \nn
\bar{\Lambda} &=& \omega \h{V}_3^{12} + \bar{\omega} \h{V}_3^{21} ,
\end{eqnarray}
and $C=\h{V}_3^{12} + \h{V}_3^{21}$ is equivalent to
$L+R=1$.
Notice that 
$L^2=L$ and $R^2 = R$ follow from $L+R =1$ and $LR = RL = 0$.

Finally the equation $L = CRC$
follows from the property of the
original vertex $CV^{12} = V^{21}C$,
since $C$ and $T$ which appear in the 
Bogoliubov transformation (\ref{Bogo})
commute with $C$.
$CV^{12} = V^{21}C$ describes a property of the vertex under
the string worldsheet orientation reversal.

\section{String Field Algebra}
In this section, we discuss the structure of the string field algebra 
${\cal A}_{str}$. 
The results in this section were essentially obtained in
\cite{KawanoOkuyama} 
using a different representation of the vertex. 

Strictly speaking, ${\cal A}_{str}$ in this section
should be understood as the zero-momentum subalgebra of string field algebra,
since the sliver state does not carry momentum. 
However, the most of the following discussions can be applied to the
whole algebra by replacing the sliver state by the $D$-instanton state,
since the Neumann coefficients including the zero-modes satisfy the same
relations as those without zero-modes
\cite{KosteleckyPotting,RSZ1,GrossTaylor}.

\subsection{String Field Oscillators}
First, we construct the string field oscillators 
$A_{\al},A^{\dag}_{\al}$
\cite{KawanoOkuyama} which satisfy the canonical commutation relations
\begin{equation}
 [A_{\al},A^{\dag}_{\bt}]_{\st}=\cob_{\al\bt}|I\ket.
\label{CCRstar}
\end{equation}
To find the explicit expression of the string field oscillators, 
it is convenient to introduce the coherent states
\begin{eqnarray}
 |z_L,z_R\ket&=&\exp(z_L\sd_L+z_R\sd_R)|\Xi\ket, \\
 I(z_L,z_R)&=& \exp(z_L\sd_L+z_R\sd_R)|I\ket.
\end{eqnarray}
The multiplication rules of these coherent states are found to be
\begin{eqnarray}
 |z_L,z_R\ket\st|w_L,w_R\ket&=&e^{z_Rw_L}|z_L,w_R\ket, \\
 I(z_L,z_R)\st |w_L,w_R\ket&=&e^{z_Rw_L}|z_L+w_L,w_R\ket, 
\label{Icoh}\\
 |z_L,z_R\ket\st I(w_L,w_R)&=&e^{z_Rw_L}|z_L,z_R+w_R\ket, 
\label{cohI}\\
 I(z_L,z_R)\st I(w_L,w_R)&=&e^{z_Rw_L}I(z_L+w_L,z_R+w_R).
\label{Izstar}
\end{eqnarray}
Using (\ref{Izstar}), we can check that the string fields
\begin{equation}
A_{\al}=\sd_{R\al}|I\ket,\quad A^{\dag}_{\al}=\sd_{L\al}|I\ket,
\label{strosci}
\end{equation}
satisfy the relation (\ref{CCRstar}).
Since the sliver state is annihilated by the string field oscillators
\begin{equation}
 A_{\al}\st|\Xi\ket=|\Xi\ket\st A^{\dag}_{\al}=0,
\end{equation}
the sliver can be identified with a projection onto the Fock vacuum 
$|0\kket$ of the string field oscillators:
\begin{equation}
 |\Xi\ket=|0\kket\bbra 0|.
\end{equation}
This identification is justified by the fact that the sliver
is naturally factorized into the vacuum of $s_L$ and $s_R$: 
\begin{equation}
 |\Xi\ket=|\Xi_L\ket\tens|\Xi_R\ket
\end{equation}
where
\begin{equation}
 s_{L\al}|\Xi_L\ket=0, \quad s_{R\al}|\Xi_R\ket=0.
\end{equation}

\subsection{Matrix Representation of String Fields}
In this subsection we construct a mapping between open string fields
and matrices. Since the identity string field has the form
$|I\ket=e^{\sd_L\sd_R}|\Xi\ket$, the string field oscillators (\ref{strosci})
can be rewritten as
\begin{eqnarray}
 &&A_{\al}=s_{L\al}|I\ket=\sd_{R\al}|I\ket, \\
 &&A^{\dag}_{\al}=\sd_{L\al}|I\ket=s_{R\al}|I\ket.
\end{eqnarray}
Using (\ref{Izstar}) recursively, 
we can show the following relations:
\begin{eqnarray}
 A_{\al_1}\st\cdots\st A_{\al_k}&=&s_{L\al_1}\cdots s_{L\al_k}|I\ket
= \sd_{R\al_1}\cdots \sd_{R\al_k}|I\ket, \\
A_{\al_1}^{\dag}\st\cdots\st A^{\dag}_{\al_k}
&=&\sd_{L\al_1}\cdots \sd_{L\al_k}|I\ket 
=s_{R\al_1}\cdots s_{R\al_k}|I\ket, 
\label{morph}
\\
{}[A_{\al},A^{\dag}_{\bt}]_{\st} &=&[s_{L\al},\sd_{L\bt}]|I\ket
= [s_{R\al},\sd_{R\bt}]|I\ket.
\end{eqnarray}
Therefore, the identity string field defines a morphism 
between worldsheet oscillators $s_{L,R}$ and the string field oscillators.
This induces the following isomorphism between the Fock spaces: 
\begin{equation}
 {\cal H}_L \cong {\cal H}_{str},\quad {\cal H}_R \cong {\cal H}_{str}^*,
\end{equation}
where ${\cal H}_{str}$ is the Fock space of string field oscillators.
Finally we have the identification between open string fields and
the matrices acting on ${\cal H}_{str}$:
\begin{equation}
 {\cal A}_{str}={\cal H}_L\tens {\cal H}_R = {\cal H}_{str}\tens
{\cal H}_{str}^*={\rm End}({\cal H}_{str}).
\end{equation}

We can construct this mapping in a more explicit way.
Using the morphism (\ref{morph}),
the star exponential of string field oscillators
turns out to be
\begin{equation}
 e_{\st}^{z_LA^{\dag}}=e^{z_L\sd_L}|I\ket,\quad 
e_{\st}^{z_RA}=e^{z_R\sd_R}|I\ket.
\label{starexp}
\end{equation}
From (\ref{Icoh}), (\ref{cohI}) and (\ref{starexp}),
we can see that the coherent states of $s_{L,R}$ and $A_{\al}$
are related as
\begin{equation}
 |z_L,z_R\ket= e^{z_L\sd_L}|I\ket\st|\Xi\ket\st e^{z_R\sd_R}|I\ket
=e^{z_LA^{\dag}}|0\kket\bbra0|e^{z_RA}=|z_L\kket\bbra z_R|.
\label{cohsA}
\end{equation}
Note that this relation is consistent with the trace 
\begin{equation}
 \Tr \Big(|z_L,z_R\ket\Big)=\bra I|z_L,z_R\ket=e^{z_Lz_R}.
\end{equation}

By expanding (\ref{cohsA}) in terms of $z_{L,R}$,
we find that the occupation number state of 
$s_{L,R}$ corresponds to the matrix element between the number states
of string field oscillators:
\begin{equation}
 |n,m\ket=\prod_{\al,\bt}{s_{L\al}^{\dag n_{\al}}s_{R\bt}^{\dag m_{\bt}}\o
\rt{n_{\al}!m_{\bt}!}}|\Xi\ket = \prod_{\al}{A_{\al}^{\dag n_{\al}}
\o\rt{n_{\al}!}}|0\kket\bbra 0|\prod_{\bt}{A_{\al}^{m_{\bt}}
\o\rt{m_{\bt}!}}=|n\kket\bbra m|.
\end{equation}
We can see that the number states behave exactly as matrix elements:
\begin{eqnarray}
 |n\kket\bbra m|\st|k\kket\bbra l|&=&\cob_{m,k}|n\kket\bbra l|, \\
 \Tr\Big(|n\kket\bbra m|\Big)&=&\cob_{n,m}.
\end{eqnarray}

Now we can construct a mapping between string fields and matrices
as follows.
Since every state can be expanded in terms of the number basis of
$s_{L,R}$,  every string field can be written as a matrix:
\begin{equation}
 |\Psi\ket=\sum_{n,m}\Psi_{n,m}|n,m\ket=\sum_{n,m}\Psi_{n,m}|n\kket\bbra m|.
\end{equation}
Especially, the identity string field corresponds to the identity matrix
\begin{equation}
 |I\ket=e^{\sd_L\sd_R}|\Xi\ket=\sum_n|n,n\ket=\sum_n|n\kket\bbra n|.
\end{equation}
Note that the 3-string vertex can also be written in terms of the number basis
\begin{equation}
 |V_3\ket_{123}=\sum_{l,m,n}|l\kket\bbra m|_1\tens |m\kket\bbra n|_2\tens
|n\kket\bbra l|_3.
\end{equation}
We can see that this vertex represents the matrix multiplication.

\section{Wedge States and $N$-string Vertices}
In this section, we first construct oscillator 
representations of wedge states, which were originally
defined as a class of surface states in \cite{RZ}. 
Then we present a closed form of
the Neumann coefficients in the $N$-string vertex and 
discuss their relation to the wedge states.

\subsection{Oscillator Representation of Wedge States}
We define the $n^{\rm th}$ wedge state 
$|n\ket_w $ by \cite{RZ}
\begin{equation}
 |n\ket_w \equiv \Big(|0\ket\Big)_{\st}^{n-1}.
\end{equation}
In this subsection, we will find an oscillator representation of the 
wedge state under the following ansatz
 \footnote{It was informed to us by B. Zwiebach that the 
oscillator representation of wedge states has been obtained 
independently by A. Sen.}
\begin{equation}
|n\ket_w={\cal N}_n^{26}\exp\lf(-\hf \a CT_n\a\ri)|0\ket.
\end{equation}
The recursion relations for ${\cal N}_n$ and $T_n$ 
follow from the relation $|n+1\ket_w=|n\ket_w\st |0\ket$:
\begin{eqnarray}
T_{n+1}&=&{X(1-T_n)\o 1-XT_n}, \\
{\cal N}_{n+1}&=&{\cal N}_n\N_{V_3}\det(1-XT_n)^{-\hf}. 
\end{eqnarray}
In the above we have assumed that $T_n$ commutes with
$X,Y$ and $Z$ which will appear to hold consistently 
when the explicit form of $T_n$ is obtained.
In order for the wedge state to converge in the
limit $n\riya\infty$, we should set the normalization ${\cal N}_{V_3}$
of the 3-string vertex to be
\begin{equation}
 \N_{V_3}=\det(1-XT)^{\hf}.
\end{equation}
By rewriting the recursion relations using (\ref{eqT})
\begin{eqnarray}
{1\o T_{n+1}-T}-{T\o 1-T^2}&=&-{1\o T}\lf({1\o T_{n}-T}-{T\o 1-T^2}\ri), \\
{{\cal N}_{n+1}\o {\cal N}_n}&=&\det\lf(-{1\o T}
{T_{n+1}-T\o T_n-T}\ri)^{\hf}, 
\end{eqnarray}
and putting the initial value $T_2=0,~{\cal N}_2=1$, 
we find 
\begin{eqnarray}
T_n&=&{T+(-T)^{n-1}\o 1-(-T)^{n}}, \label{TN}\\
{\cal N}_n&=&\det\lf[{1-T^2\o 1-(-T)^{n}}\ri]^{\hf}. 
\end{eqnarray}
Note that the normalization of the identity $\N_1$ and the 3-string vertex
$\N_{V_3}$ satisfy the relation found in \cite{GrossTaylor}:
\begin{equation}
 \N_1\N_{V_3}=\det(1-X)^{\hf}.
\end{equation}
Note also that $\N_{V_3}$ is equal to $\N_3$. 
The final expression of the wedge state is
\begin{equation}
|n\ket_w=\det\lf[{1-T^2\o 1-(-T)^{n}}\ri]^{{26\o2}}
\exp\lf(-\hf\a C{T+(-T)^{n-1}\o 1-(-T)^{n}}\a\ri)|0\ket. 
\end{equation}
In this normalization, the sliver state which
can be defined as a limit of the wedge states \cite{RZ} becomes
\begin{equation}
|\Xi\ket=\lim_{n\riya\infty}|n\ket_w=\det(1-T^2)^{{26\o2}}
\exp\lf(-\hf\a CT\a\ri)|0\ket .
\end{equation}
Here we assumed
\begin{equation}
 \label{limT}
\lim_{n\riya\infty}T^{n}=0. 
\end{equation}
Although we do not have a  proof of this equation, 
it was discussed in \cite{KosteleckyPotting} that the absolute values 
of the eigenvalues of $T$ are less than one. 
The existence of the smooth limit to 
the sliver state \cite{RSZ1} also supports the existence of this limit,
since the above discussion gives
an analytical explanation for
the equivalence
between two definitions of the sliver,
namely the original definition in \cite{RZ} and
the oscillator representation \cite{KosteleckyPotting}.

Up to a normalization, the wedge state over the sliver state is given by
\begin{equation}
|n\ket_w=\exp\lf(-\hf\sd C(-T)^{n-1}\sd\ri)|\Xi\ket 
\end{equation}

\subsection{Neumann Coefficients of $N$-string Vertex}
In this subsection, we will present a closed formula 
of the Neumann coefficients $V^{rs}_N$ in the $N$-string vertex
\begin{equation}
|V_N\ket=\exp\lf(-\hf \sum_{r,s=1}^N\a_rV^{rs}_N\a_s\ri)|0\ket_{1\cdots N}.
\end{equation}
See \cite{CommaN} for the earlier discussion on the relation between 
the comma vertex and the Neumann coefficient.

We can easily generalize the comma representation of the 3-string vertex 
(\ref{comma3})
to the  $N$-string  vertex:
\begin{equation}
 |V_N\ket=\exp\lf(-\sum_{r=1}^N\sd_rCL\sd_{r+1}\ri)|\Xi\ket_{1\cdots N}=
\exp\lf(-\hf \sum_{r,s=1}^N\sd_r\h{V}^{rs}_N
\sd_s\ri)|\Xi\ket_{1\cdots N},
\end{equation}
where $\h{V}_N^{rs}$ is given by
\begin{equation}
C\h{V}_N=\lf(
\begin{array}{cccccc}
0&L&0&0&\cdots& R\\
R&0&L&0&\cdots&0 \\
0&R&0&L&\cdots&0 \\
 & &\ddots&\ddots &\ddots& \\
0&\cdots&0& R&0&L\\
L&0&\cdots&0&R&0 
\end{array}\ri),
\end{equation}
in other words,
\begin{equation}
C\h{V}^{rs}_N=\cob_{s-r,1}L+\cob_{r-s,1}R.
\end{equation}
This expression of $N$-string vertex satisfies the 
required gluing relations \cite{KawanoOkuyama}
\begin{eqnarray}
&& {}_N\bra I|V_N\ket_{1\cdots N}=|V_{N-1}\ket_{1\cdots N-1}, 
\label{IVN}\\
 &&\bra R(K,L)||V_{M+1}\ket_{1\cdots M,K}|V_{N+1}\ket_{L,M+1,\cdots,M+N}
= |V_{M+N}\ket_{1\cdots M+N}.
\end{eqnarray}
By going back to the original variables, we find that 
$V^{rs}_N$ is given by
\begin{equation}
CV_N=(T+C\h{V}_N)(1+C\h{V}_NT)^{-1} ,
\label{VhatV}
\end{equation}
where $(T+C\h{V}_N)^{rs}=T\cob^{rs}+C\h{V}_N^{rs}$.
For example, 4-string coefficient $V_4^{rs}$ becomes
\begin{eqnarray}
&&CV_4^{11}=-CV^{13}_4={T\o 1+T^2}={X\o 1+X}, \nn
&&CV_4^{12}={L+T^2R\o 1+T^2}=\hf\lf(1+{Y-Z\o 1+X}\ri), \\ 
&&CV_4^{14}={R+T^2L\o 1+T^2}=\hf\lf(1+{Z-Y\o 1+X}\ri). \nonumber
\end{eqnarray}
This expression agrees with the result in \cite{GrossJevicki}. 

For general $N$, we can find 
the Neumann coefficients as follows.
Since $V_N^{rs}$ depends only on $s-r~{\rm mod}~N$,
we can regard it as a vector.
For the object $A_s$ with a mod $N$ index,
we define the discrete Fourier transformation
\begin{equation}
A_s={1\o N}\sum_{k=1}^N\om^{ks}\h{A}_k ,
\end{equation}
where $\om=\exp(2\pi i/N)$.
Then the convolution of two objects $A_r$ and $B_s$ becomes
\begin{equation}
\sum_{r=1}^{N}A_rB_{s-r}={1\o N}\sum_{k=1}^N\om^{ks}\h{A}_k\h{B}_k.
\end{equation}
Using the Fourier representation of the matrices in the right-hand-side of
(\ref{VhatV})
\begin{eqnarray}
 (T+C\h{V}_N)^{rs}&=&{1\o N}\sum_{k=1}^N\om^{k(s-r)}
\Big[(T+\om^{-k})L+(T+\om^k)R\Big], \\
\Big[(1+C\h{V}_NT)^{-1}\Big]^{rs}&=&{1\o N}\sum_{k=1}^N\om^{k(s-r)}
\lf[{1\o 1+T\om^{-k}}L+{1\o 1+T\om^{k}}R\ri],
\end{eqnarray}
the Neumann coefficients for the general $N$ are found to be
\begin{eqnarray}
CV_N^{rs}&=&{1\o N}\sum_{k=1}^N\om^{k(s-r)}
\lf[{T+\om^{-k}\o 1+T\om^{-k}}L+
{T+\om^k\o 1+T\om^k}R\ri] \nn
&=&{T(-T)^{[s-r]}+(-T)^{[s-r-1]}\o 1-(-T)^N}L
+{T(-T)^{[r-s]}+(-T)^{[r-s-1]}\o 1-(-T)^N}R,
\label{VNclose}
\end{eqnarray}
where  
\begin{equation}
[s]\equiv s~{\rm mod}~N,\quad [s]\in\{0,1,\cdots, N-1\}. 
\end{equation}
One can check that this form of $V^{rs}_N$ satisfies the descent relation
\begin{equation}
CV_{N-1}^{rs}=CV_N^{rs}+{CV^{rN}_NCV_N^{Ns}\o 1-CV_N^{11}},
\end{equation}
which comes from (\ref{IVN}).
One can also check that
\begin{equation}
 CV_N^{rs}=CV_{2N}^{rs}+CV_{2N}^{r,s+N}
=CV_{3N}^{rs}+CV_{3N}^{r,s+N}+CV_{3N}^{r,s+2N}
=\cdots=\sum_{l=0}^{k-1}CV_{kN}^{r,s+lN}.
\end{equation}
This is a generalization of the well-known relation between
3- and 6-string vertices \cite{GrossJevicki}.

Here we comment on the relation between the $N$-string 
vertices and the wedge states.
The integral representation of Neumann coefficient is given by
\begin{equation}
(V_N^{rs})_{nm}=-{1\o\rt{nm}}\oint {dz\o2\pi i}\oint {dw\o 2\pi i}
z^{-n}w^{-m}{\del_z f^{(r)}_N(z)\del_wf^{(s)}_N(w)
\o [f^{(r)}_N(z)-f^{(s)}_N(w)]^2}, 
\end{equation}
where
\begin{equation}
f^{(r)}_N(z)=\tan\lf({2\o N}\tan^{-1}z-{\pi r\o N}\ri)
=S^r\lf(\tan\lf({2\o N}\tan^{-1}z\ri)\ri). 
\end{equation}
$S$ is an $SL(2,\R)$ transformation
\begin{equation}
S(z)={z\cos{\pi\o N}-\sin{\pi\o N}\o z\sin{\pi\o N}+\cos{\pi\o N}},
\end{equation}
which satisfies $S^N=1$. Since the matrix $CT_N$ in the wedge state $|N\ket_w$
has the same integral representation as $V^{11}_N$, they should be the same.
This can be also understood from the defining equations
$|N\ket_w = \Big(|0\ket\Big)_{\st}^{N-1}
={}_1\bra 0| \cdots {}_{N-1}\bra 0| |V_N\ket$.
As expected, the diagonal component in (\ref{VNclose}) agrees with  (\ref{TN})
\begin{equation}
CV_N^{11}=T_N={T+(-T)^{N-1}\o 1-(-T)^N}.
\end{equation}

\section{Summary and Discussions}
In this paper, we analyzed the structure of the string field algebra
using the so-called comma vertex. We obtained a closed formula of the
$N$-string Neumann coefficient and showed that the diagonal element of
this coefficient agrees with the width matrix of
the wedge state.

It will be interesting to extend the idea of the comma vertex to
the general CFT background. In fact, the 3-string vertex over the sliver
state was constructed in an arbitrary CFT \cite{David}. Up to now,
its relation to the comma vertex is unclear.
It is also important to study the comma representation
in the ghost sector. In \cite{KosteleckyPotting}, it was argued that
the 3-string vertex in the ghost sector can also be written in a comma form
except for the subtlety associated with the ghost zero-modes.

\vskip 10mm
\centerline{\bf Acknowledgments}
We would like to thank N. Ishibashi,
T. Kawano, T. Kugo, and B. Zwiebach for useful discussions.
K.$\,$O. was supported in part by JSPS Research Fellowships for Young
Scientists.

\end{document}